\documentstyle[12pt]{article}

\begin{document}

\author{Amir H. Abbassi \\
Department of Physics, School of Sciences,\\
Tarbiat Modarres University,P.O.Box 14155-4838,\\
Tehran, Iran \\
E-Mail: ahabbasi@net1cs.modares.ac.ir}
\title{$An\;Extension\;of\;Schwarzschild\;Space\;$\\
$to\;\;r=0$}
\date{July,2001 }
\maketitle

\begin{abstract}
A more rigorous treatment of the Schwarzschild metric by making use of the
energy-momentum tensor of a single point particle  as source term shows that

\begin{center}
$g_{00}=-\{1-\frac{2GM}{c^2r}-\frac{8G^2 M^2}{c^4 r^2}(\theta (r)-1)\}\exp [2(\theta
(r)-1)]$

$g_{rr}=\{1-\frac{2GM}{c^2r}-\frac{8G^2 M^2}{c^4 r^2}(\theta (r)-1)\}^{-1}$
\end{center}

\noindent The existence of a discontinuity at $r=0$ leads to an infinite
repulsive force that will change the ultimate fate of a free fall test
particle to a bouncing state.

\bigskip\ 

\end{abstract}

\newpage\ We intend to show that at $r=0$ there is a mathematical
inconsistency in derivation of the Schwarzschild metric which is the static
solution of Einstein field equations for vacuum space around a point mass
M[1] . It is a simple case that is propounded in any related text in GR.
Usually similar approaches to the problem are followed[2-6]. We use the
notations and derivations of Weinberg[2] for the sake of definiteness.

\noindent The standard form for general isotropic metric is :

\begin{equation}
d\tau ^2=B(r)dt^2-A(r)dr^2-r^2d\theta ^2-r^2\sin ^2\theta d\varphi ^2
\end{equation} \label{1}

\noindent The components of Ricci tensor for the metric (1) are:

\begin{equation}
R_{rr}=\frac{B^{^{\prime \prime }}}{2B}-\frac{B^{^{\prime }}}{4B}(\frac{%
A^{^{\prime }}}A+\frac{B^{^{\prime }}}B)-\frac{A^{^{\prime }}}{rA}
\end{equation} \label{2}

\begin{equation}
R_{\theta \theta }=-1+\frac r{2A}(-\frac{A^{^{\prime }}}A+\frac{B^{^{\prime
}}}B)+\frac 1A
\end{equation}\label{3}

\begin{equation}
R_{\varphi \varphi }=\sin ^2\theta \;R_{\theta \theta }
\end{equation}\label{4}

\begin{equation}
R_{tt}=-\frac{B^{^{\prime \prime }}}{2A}+\frac{B^{^{\prime }}}{4A}(\frac{%
A^{^{\prime }}}A+\frac{B^{^{\prime }}}B)-\frac{B^{^{\prime }}}{rA}
\end{equation}\label{5}

\[
R_{\mu \nu }=0\;\;\;\;\;for\;\mu \neq \nu 
\]
Prime means differentiation with respect to r. The applied field equation
for empty space is

\begin{equation}
R_{\mu \nu }=0
\end{equation}\label{6}
Using (6) we may conclude that $\frac{R_{rr}}A+\frac{R_{tt}}B=0$ , and by
inserting (2) and (5) in this relation we get

\begin{equation}
-\frac 1{rA}(\frac{A^{^{\prime }}}A+\frac{B^{^{\prime }}}B)=0
\end{equation}\label{7}
Integration of (7) with respect to r and imposing Minkowski metric as r
tends to infinity for boundary condition gives

\begin{equation}
A=\frac 1B
\end{equation}\label{8}
Now by arranging (2) ,(3), (6) and (8) we obtain

\begin{equation}
R_{\theta \theta }=-1+rB^{^{\prime }}+B=0
\end{equation}\label{9}

\begin{equation}
R_{rr}=\frac{B^{^{\prime \prime }}}{2B}+\frac{B^{^{\prime }}}{rB}=0
\end{equation}\label{10}
Integrating (9) with respect to r gives

\begin{equation}
rB=r+const.
\end{equation}\label{11}
For fixing the constant of integration in (11) we recall that at great
distances from mass M,the component of $g_{tt}=-B$ must approach to $-1-%
\frac{2\Phi }{c^2}$ , where $\Phi \;$is the Newtonian potential $-\frac{GM}r$%
. Hence it is equal to $-\frac{2GM}{c^2}$ and our final result is

\begin{equation}
B=1-\frac{2GM}{c^2r}
\end{equation}\label{12}

According to the correspondence principle general relativity must agree on
the one hand with special relativity in the absence of gravitation and on
the other hand with Newtonian theory of gravitation in the limit of weak
gravitational fields and low velocities. The corresponding field equation of
this problem in Newtonian gravity is the Poisson equation

\[
\nabla ^2\Phi =4\pi GM\delta (\stackrel{\rightarrow }{r}) 
\]
It is remarkable to notice that this equation is defined in the whole space
even at $r=0$ and its solution which is proportional to inverse of $r$
satisfies the field everywhere even at $r=0$. Thus we have right to expect
that the general relativity field equations which are supposed to be yielded
to this equation at weak field limit must be well defined at $r=0$ [7,8].

Since the singularity at $r=0$ is not a hypothetical case and it is the inevitable end state of gravitational collapse of a massive body , the external gravitational field of our point particle must somehow be traced to the source located at $r=0$. We consider this classical problem not in the belief that it is 
the final word , but rather a useful start which quantum
phenomena will no doubt modify. It can then be used as
a background for the more advanced theories. Such advanced
theories are important , but they have achieved their greatest
successes when the corresponding classical problem has been 
fully understood.

The important point is that for this case $T_{\mu \nu }$ does not vanish
everywhere, it is nonzero at $r$ exactly equal to zero. For finding the $%
T_{\mu \nu }$ of a mass point M located at $x_1$ we may use the action [9]

\begin{equation}
I_M=-M\int_{-\infty }^{+\infty }dp[-g_{\mu \nu }(x_1)\frac{dx_1^\mu (p)}{dp}%
\frac{dx_1^\nu (p)}{dp}]^{\frac 12}
\end{equation}\label{13}
$T^{\mu \nu }$ is defined to be

\begin{equation}
\delta I_M=\frac 12\int d^4x\sqrt{g(x)}T^{\mu \nu }(x)\delta g_{\mu \nu }(x)
\end{equation}\label{14}
A straight forward calculation gives

\begin{equation}
\delta I_M=\frac 12M\int_{-\infty }^{+\infty }dp[-g_{\mu \nu }(x_1(p))\frac{%
dx_1^\mu (p)}{dp}\frac{dx_1^\nu (p)}{dp}]^{-\frac 12}\frac{dx_1^\lambda (p)}{%
dp}\frac{dx_1^\kappa (p)}{dp}\delta g_{\lambda \kappa }(x_1(p))
\end{equation}\label{15}
This is of the form (14) with

\begin{equation}
T^{\lambda \kappa }(x)=g^{-\frac 12}(x)M\int_{-\infty }^{+\infty }d\tau _1%
\frac{dx_1^\lambda }{d\tau _1}\frac{dx_1^\kappa }{d\tau _1}\delta ^4(x-x_1)
\end{equation}\label{16}
if we take $ \int d^4x \delta ^4 (x-x_1)=1$ as is taken by Weinberg. It would be more concrete if we take 
$\int \sqrt{g} d^4 x \delta ^4 (x-x_1)=1$ which in this case we come to 
\begin{equation}
T^{\lambda \kappa}(x) = M\int_{-\infty}^{+\infty}dp
[-g_{\mu\nu}(x_1)\frac{dx_1^\mu}{dp}\frac{dx_1^\nu}
{dp}]^{-\frac 12}\frac{dx_1^\lambda (p)}{dp}
\frac{dx_1^\kappa (p)}{dp} \delta ^4 (x-x_1 (p))
\end{equation}\label{17}

For a partcile located at the origin we have $x_1^\mu =(x_1^0 , 0)$ , and in  
 a frame at rest with respect to the particle the components of the
four velocity are

\begin{equation}
u^i=0\;,\;u^t=B^{-\frac 12}\;,\;u_t=-B^{\frac 12}
\end{equation}\label{18}

Thus the (00)-component of (17) is
\begin{eqnarray}
T^{00}(x)&=& M\int_{-\infty}^{+\infty} dp [B(0)]^{-\frac12}
\frac{dx_1^0}{dp} \delta^4 (x-x_1) \nonumber\\
&=&M\int_{-\infty}^{+\infty}dx_1^0 \frac{\delta^4 (x-x_1)}
{\sqrt{B(0)}}
\end{eqnarray}\label{19}
To find out the form of $\delta^4 (x-x_1)$ it is more 
convenient to work in the coordinate system
$x^\mu =(x^0 , r, q= cos\theta , \varphi)$.
In this system we have $g_{qq}=\frac{g_{\theta\theta}}{sin^2\theta}$
and $\sqrt g=r^2 \sqrt{AB}$. Thus we may write
\begin{equation}
\int r^2 \sqrt{AB} \delta^4 (x-x_1)d^4x =1
\end{equation}\label{20}
(20) will be satisfied if we take
\begin{equation}
\delta^4(x-x_1)=\frac{\delta (x^0-x_1^0)\delta(r)}
{2\pi r^2\sqrt{AB}}
\end{equation}\label{21}
where we have $\int_{0}^{+\infty}\delta(r)dr=\frac12$.

Inserting (21) in (19) gives 
\begin{equation}
T^{00}(x)= \frac{M\delta(r)}{2\pi r^2 \sqrt{ABB(0)}}
\end{equation}\label{22}
We should notice that $ T^{tt}=c^2 T^{00}$ and 
$ T^{\tau\tau}=BT^{tt}$.  Thus we have 
\begin{equation}
T^{tt}(x)=\frac{Mc^2\delta(r)}{2\pi r^2 \sqrt{ABB(0)}}
\end{equation}\label{23}
and
\begin{equation}
T^{\tau\tau}=\frac{Mc^2 \delta(r)}{2\pi r^2}\sqrt{\frac
{B}{AB(0)}}
\end{equation}\label{24}
Using (24) we may write 
\begin{equation}
\int T^{\tau\tau}4\pi r^2 \sqrt{A} dr= M c^2
\end{equation}\label{25}
as we should have expected. Also we have

\begin{equation}
T_{00}(x)=\frac{Mc^2\delta (r)}{2\pi r^2}B\sqrt{\frac {B}{AB(0)}}
\end{equation}\label{26}
Then $T\equiv g^{\mu \nu }T_{\mu \nu }$ is

\begin{equation}
T=-\frac{Mc^2\delta (r)}{2\pi r^2}\sqrt{\frac{B}{AB(0)}}
\end{equation}\label{27}
\[
\]

Now the components of $S_{\mu \nu }$ which are defined as $S_{\mu \nu
}=T_{\mu \nu }-\frac 12g_{\mu \nu }T$ are:

\begin{equation}
S_{tt}=\frac{Mc^2\delta (r)}{4\pi r^2}B\sqrt{\frac B{AB(0)}}
\end{equation}\label{28}

\begin{equation}
S_{rr}=\frac{Mc^2\delta (r)}{4\pi r^2}\sqrt{\frac {AB}{B(0)}}
\end{equation}\label{29}

\begin{equation}
S_{\theta \theta }=\frac{Mc^2\delta (r)}{4\pi \sqrt{\frac {B}{AB(0)}}}
\end{equation}\label{30}

\begin{equation}
S_{\varphi \varphi }=S_{\theta \theta }\sin ^2\theta
\end{equation}\label{31}

\begin{equation}
S_{\mu \nu }=0\;\;\;\;\;\;\mu \neq \nu
\end{equation}\label{32}
The rigorous treatment of the problem is to solve the following equations

\begin{equation}
R_{\mu \nu }=-\frac{8\pi G}{c^4}S_{\mu \nu }
\end{equation}\label{33}
where the components of $S_{\mu \nu }$are given by (28)-(32). Using (33) and
(2)-(5) we obtain

\begin{equation}
R_{rr}=\frac{B^{^{\prime \prime }}}{2B}-\frac{B^{^{\prime }}}{4B}(\frac{%
A^{^{\prime }}}A+\frac{B^{^{\prime }}}B)-\frac{A^{^{\prime }}}{rA}=-\frac{%
2GM\delta (r)}{c^2r^2}\sqrt{\frac {AB}{B(0)}}
\end{equation}\label{34}

\begin{equation}
R_{\theta \theta }=-1+\frac r{2A}(-\frac{A^{^{\prime }}}A+\frac{B^{^{\prime
}}}B)+\frac 1A=-\frac{2GM\delta (r)}{c^2}\sqrt{\frac  {B}{AB(0)}}
\end{equation}\label{35}

\begin{equation}
R_{\varphi \varphi }=\sin ^2\theta \;R_{\theta \theta }
\end{equation}\label{36}

\begin{equation}
R_{tt}=-\frac{B^{^{\prime \prime }}}{2A}+\frac{B^{^{\prime }}}{4A}(\frac{%
A^{^{\prime }}}A+\frac{B^{^{\prime }}}B)-\frac{B^{^{\prime }}}{rA}=-\frac{%
2GM\delta (r)}{c^2r^2}B\sqrt{\frac{B}{AB(0)}}
\end{equation}\label{37}
It can be checked easily that the usual solutions $AB=1$ and $B=1-\frac{2GM}{%
c^2r}$ do not satisfy (34)-(37).
This point has been recognized by Narlikar too [10].
He says when we try to solve for a point mass gravitational field
the condition $AB=1$ yields to an inconsistency at $r=0$ because this gives ${T^0}_0={T^1}_1$ while for a static problem we need
${T^1}_1 =0$. He argues that it would be tempty to take the easy way out of the problem by taking that the Schwarzschild coordinates are in appropriate at $r=0$. \\

Using (34) and (37), $\frac{R_{rr}}A+\frac{%
R_{tt}}B$ gives

\begin{equation}
-\frac 1{rA}(\frac{A^{^{\prime }}}A+\frac{B^{^{\prime }}}B)=-\frac{4GM\delta
(r)}{c^2r^2}\sqrt{\frac{B}{AB(0)}}
\end{equation}\label{38}
and

\begin{equation}
\frac{A^{^{\prime }}}A+\frac{B^{^{\prime }}}B=\frac{4GM\delta (r)}{c^2r}%
\sqrt{\frac {AB}{B(0)}}
\end{equation}\label{39}
Now we may combine (39) and (35) to write

\begin{equation}
-1-\frac{rA^{^{\prime }}}{A^2}+\frac{2GM\delta (r)}{c^2}\sqrt{\frac B{AB(0)}}+\frac 1A=-%
\frac{2GM\delta (r)}{c^2}\sqrt{\frac B{AB(0)}}
\end{equation}\label{40}
which gives

\begin{equation}
-1+\frac d{dr}(\frac rA)+\frac{4GM\delta (r)}{c^2}\sqrt{\frac B{AB(0)}}=0
\end{equation}\label{41}
Integrating (41) with respect to $r$ from $+\infty $ to $r$ and
taking that\\
 $\sqrt{\frac B{AB(0)}}\delta(r)=\frac{\delta(r)}
{\sqrt{A(0)}}$ , yields

\begin{equation}
-r+\frac rA+\frac{4GM(\theta (r)-1)}{c^2\sqrt{A(0)}}\ =const. \;\; ,
\end{equation}\label{42}
where $\theta (x)=\left\{ 
\begin{array}{l}
1\;\;\;x>0 \\ 
\frac 12\;\;x=0 \\ 
0\;\;\;x<0
\end{array}
\right . $.

For finding the integration constant in (42) we may impose this fact that at
large distances we have $AB=1$. Then we get

\begin{equation}
-r+rB+0=const. \;\; ,
\end{equation}\label{43}\\
\noindent and at this range B is equal to $1-\frac{2GM}{c^2r}.$ By inserting this
value for B in (43) the constant in (42) and (43) will be found to be equal
to $-\frac{2GM}{c^2}$. Putting this in (42) results that

\begin{equation}
A=\frac 1{1-\frac{2GM}{c^2r}[1+\frac{\theta (r)-1}{\sqrt{A(0)}}]}
\end{equation}\label{44}
Taking the limit of $r\rightarrow 0$, (44) gives
\begin{equation}
A(0)=\lim_{r\rightarrow 0}\frac{c^2 r}{2GM}\sqrt{A(0)}
\end{equation}\label{45}
or
\begin{equation}
A(0)=\lim_{r \rightarrow 0}\left( \frac{c^2 r}{2GM}\right)^2
\end{equation}\label{46}
Considering (46), we may write (44) as 
\begin{equation}
A=\frac{1}{1-\frac{2GM}{c^2 r}-\frac{8G^2M^2}{c^4 r^2}
(\theta (r)-1)}
\end{equation}\label{47}
because the correction term is merely effective at $r=0$.
 
Then we have
\begin{equation}
\frac{A^{^{\prime }}}A=\frac{-\frac{2GM}{c^2r^2}-\frac{16G^2M^2}{c^4r^3}
(\theta (r)-1)+\frac{8G^2M^2\delta (r)}{c^2r^2}}{ 1-\frac{2GM}{c^2r}-\frac{8G^2M^2}{c^4 r^2}(\theta (r)-1)}
\end{equation}\label{48}
Putting (48) in (39) and using this fact that at r.h.s of it $\sqrt{\frac{AB}{B(0)}}
$indeed is $\sqrt{A(0)}$  and (46) yield

\begin{equation}
\frac{B^{^{\prime }}}B=\frac{\frac{2GM}{c^2r^2}+\frac{16G^2M^2}{c^4 r^3}(\theta (r)-1)}{ 1-\frac{2GM}{c^2r}-\frac{8G^2M^2}
{c^4 r^2}(\theta (r)-1)}
\end{equation}\label{49}
Integrating (49) with respect to $r$ from $+\infty $ to $r$ and imposing $%
B(+\infty )=1$ gives

\begin{equation}
B=\{1-\frac{2GM}{c^2r}-\frac{8G^2 M^2}{c^4 r^2}
(\theta (r)-1)\}\exp
[2(\theta (r)-1)]
\end{equation}\label{50}
Using (47) and (50) we have

\begin{equation}
AB=\exp [2(\theta (r)-1)]
\end{equation}\label{51}

Finally our rigorous result for Schwarzschild metric is:

\begin{eqnarray}
ds^2 &=&c^2\{1-\frac{2GM}{c^2r}-\frac{8G^2M^2}{
C^4 r^2}(\theta (r)-1)\}\exp [2(\theta
(r)-1)]dt^2  \nonumber \\
&&-\frac 1{\ 1-\frac{2GM}{c^2r}-\frac{8G^2M^2}{c^4 r^2}(\theta (r)-1)}dr^2-r^2(d\theta
^2+\sin ^2\theta \;d\varphi ^2)
\end{eqnarray}\label{52}

It can be easily verified that this metric satisfies the field equation
exactly at each point even at $r=0.$ In a mathematical rigorous treatment
one may truly expect that the theory of distribution should be considered
[11-15]. This is not a possible task at this stage because distribution in
curved space-time is not well known. Here theta and delta functions have been
considered in the same manner which are commonly being used in physics
literature. This work is a step toward a perfect treatment in this sense.
Another point to notice is that in Newtonian gravity the potential is a
continuous function even at $r=0$ and the force is always attractive. In
this result there exist a discontinuity in the metric at $r=0$. The
neighborhood of $r=0$ that is where $r<<2M$ is a domain in which the field
is very strong and therefore there is no weak field limit corresponding to
it. This phenomenon is a pure general relativistic effect which has no
analog in Newtonian gravity. The existence of this discontinuity means that
we have an infinite repulsive force at $r=0$ (see appendix). 
For more clarification we will find the tidal forces and the
scalar invariant ${R^\mu}_{\nu\lambda\rho}
{R_\mu}^{\nu\lambda\rho}$ for this metric in the 
next section. This repulsive
force causes to change the ultimate fate of a free fall test particle from a
collapse into intrinsic singularity to a bouncing state. The prediction of
this phenomenon itself may be considered as a justification for this
calculations which was rather detailed and the presented treatment for
determining the form of the line element.

\section{Tidal Forces}

The tidal effect results in an elongation of the distribution 
in the direction of motion and a compression of the distribution
in transverse directions.  The same effect occurs in a body falling
towards a spherical object in general relativity.  We can
gain some idea of this by considering the equation of geodesic
deviation in the form [16]
\begin{equation}
\frac{D^2 \eta^{\alpha}}{D\tau^2}-{R^a}_{bcd}
{e^{ \alpha}}_a {\it v^b}{\it v^c}{e_{\beta}}^d
\eta^{ \beta}=0 
\end{equation}\label{53}\\
\noindent for the spacelike components of the orthogonal
connecting vector $\eta^a$ connecting two neighbouring 
particles in free fall.\\
Let the frame ${e_i}^a$ be defined in Schwarzschild coordinate 
as 
\begin{equation}
\begin{array}{lll}
{e_0}^a & = & B^{-\frac 12} (1,0,0,0)\\
{e_1}^a & = & A{-\frac 12} (0,1,0,0)\\
{e_2}^a & = & r^{-1}(0,0,1,0)\\
{e_3}^a & = & (r sin\theta)^{-1}(0,0,0,1)
\end{array}
\end{equation}\label{54}\\
\noindent and denote the components $\eta^{\alpha}$
by 
\begin{equation}
\eta^{\alpha}= (\eta^1,\eta^2,\eta^3)
=(\eta^r, \eta^\theta, \eta^\varphi)
\end{equation}\label{55}\\

Using (1) the nonvanishing components of the curvature
tensor are :
\begin{equation}
\begin{array}{lllll}
{R^t}_{rtr}&=-\frac{B^{^{\prime\prime}}}{2B}\quad\quad&, \quad\quad&{R^r}_{trt}&=\frac{B^{^{\prime\prime}}}{2A}\\

{R^t}_{\theta t\theta}&=\frac{rA^{^\prime}}{2A^2}\quad\quad&,\quad \quad&{R^r}_{\theta r \theta}&={R^t}_{\theta t \theta}\\

{R^t}_{\varphi t\varphi}& ={R^t}_{\theta t\theta} \;Sin^2\theta\quad\quad&,\quad\quad&{R^r}_{\varphi r \varphi}& =
{R^t}_{\varphi t\varphi}\\
{R^\theta}_{t\theta t}&=\frac{B^{^\prime}}{2rA}\quad
\quad&,\quad\quad
&{R^\varphi}_{t\varphi t}&={R^\theta}_{t\theta t}\\
{R^\theta}_{r\theta r}&=\frac{A^{^\prime}}{2rA}\quad
\quad&,\quad\quad&{R^\varphi}_{r\varphi r}&=
{R^\theta}_{r\theta r}\\
{R^\theta}_{\varphi\theta\varphi}&={R^\varphi}_
{\theta\varphi\theta}\; Sin^2\theta \quad\quad&,
\quad\quad&{R^\varphi}_{\theta\varphi\theta}&=
1-\frac 1A
\end{array}
\end{equation}\label{56}\\
\noindent By using (56) and (54) , (53) gives the following
\begin{equation}
\frac{D^2\eta^r}{D\tau^2}+\frac{B^{^{\prime\prime}}}
{2AB}\eta^r =0
\end{equation}\label{57}

\begin{equation}
\frac{D^2\eta^\theta}{D\tau^2}+\frac{B^{^\prime}}{2rAB}
\eta^\theta =0
\end{equation}\label{58}

\begin{equation}
\frac{D^2\eta^\varphi}{D\tau^2}+\frac{B^{^\prime}}
{2rAB}\eta^\varphi =0
\end{equation}\label{59}
Using (47), (50) and(51) we have 
\begin{equation}
\frac{B^{^\prime}}{2AB}=\frac{GM}{c^2r^2}+\frac
{8G^2M^2}{c^4r^3}(\theta (r)-1)-\frac{4G^2M^2}
{c^4r^2}\delta (r) +\frac{\delta (r)}{A}
\end{equation}\label{60}
By (46) we have, $\frac{\delta (r)}{A}=\frac{4G^2M^2}{c^4r^2}
\delta (r)$ , then (60) becomes 
\begin{equation}
\frac{B^{^\prime}}{2AB}=\frac{GM}{c^2r^2}+
\frac{8G^2M^2}{c^4r^3}(\theta (r)-1)
\end{equation}\label{61}
Next (34) gives 
\begin{equation}
\frac{B^{^{\prime\prime}}}{2AB}=
\frac{B^{^\prime}}{4AB}(\frac{A^{^\prime}}A
+\frac{B^{^\prime}}B)+ \frac{A^{^\prime}}{rA^2}
-\frac{2GM}{c^2r}\sqrt{\frac B{AB(0)}} \delta (r)
\end{equation}\label{62}
Inserting (61) , (39) and (46) in (62) yields
\begin{equation}
\frac{B^{^{\prime\prime}}}{2AB}=
-\frac{2GM}{c^2r^3}-\frac{16G^2M^2}{c^4r^4}
(\theta (r)-1)+\frac{GM}{c^2r^2}\delta (r)
\end{equation}\label{63}
Finally putting (61) and (63) in (57)-(59) leads to
\begin{eqnarray}
\frac{D^2\eta^r}{D\tau^2}&=&
\{\frac{2GM}{c^2r^3}+\frac{16G^2M^2}{c^4r^4}
(\theta (r)-1)-\frac{GM}{c^2r^2}\delta (r)\}\eta^r \\
\frac{D^2\eta^\theta}{D\tau^2}&=&
\{-\frac{GM}{c^2r^3}-\frac{8G^2M^2}{c^4r^4}
(\theta (r)-1)\}\eta^\theta \\
\frac{D^2\eta^\varphi}{D\tau^2}&=&
\{-\frac{GM}{c^2r^3}-\frac{8G^2M^2}{c^4r^4}
(\theta (r)-1)\}\eta^\varphi
\end{eqnarray}\label{66}
(64)-(66) exhibit the usual tidal force of the
Schwarzschild metric for $r\neq 0$. At $r=0$ the correction
terms are dominant and have opposite sign to the 
first terms.

\section{Scalar Invariant}

Since the metric components have no cross terms and 
furthermore the curvature tensors have some algebric
properties, the Riemann tensor scalar invariant take
a simple form as 
\begin{equation}
{R^a}_{bcd}{R_a}^{bcd}=2\sum_{a\neq b}
(g^aa {R^b}_{aba})^2
\end{equation}\label{67}
By using (56) we may write (67) as 
\begin{equation}
{R^a}_{bcd}{R_a}^{bcd}=(\frac{B^{^{\prime\prime}}}{AB})^2
+4(\frac{A^{^\prime}}{rA^2})^2 +\frac 4{r^4}
(1-\frac 1A)^2
\end{equation}\label{68}
Inserting (63) and (47) in (68) gives
\begin{eqnarray}
{R^a}_{bcd}{R_a}^{bcd}=4\{-\frac{2GM}{c^2r^3}-\frac{16G^2M^2}{c^4r^4}
(\theta (r)-1)+\frac{GM}{c^2r^2}\delta (r)\}^2\nonumber\\
+\frac 4{r^2}\{\frac{2GM}{c^2r^2}+\frac{16G^2M^2}
{c^4r^3}(\theta (r)-1)-\frac{8G^2M^2}{c^4r^2}\delta (r)\}^2
\nonumber\\
+\frac 4{r^2}\{\frac{2GM}{c^2r^2}+\frac{16G^2M^2}
{c^4r^3}(\theta (r)-1)\}^2
\end{eqnarray}\label{69}
(69) shows that ${R^a}_{bcd}{R_a}^{bcd}$ behaves like
$\frac{48G^2M^2}{c^4r^6}$ for $r\neq 0$, the same result
of the Schwaezschild metric. At $r=0$ in addition of this term
there are some correction terms which all of them diverge.
 
\section{Remark}

According to singularity theorems a condition for the existence of the
singularity is that $R_{ab}K^aK^b\geq 0$ for every non-spacelike vector $K.$
Now using the four velocity (18) as a vector we obtain in this case

\begin{equation}
R_{ab}u^au^b=-\frac{GM\delta (r)}{r^2}\sqrt{\frac B{AB(0)}}
=-\frac{\delta (r)}{2r}
\end{equation}\label{70}\\
\noindent which is negative for $r=0.$ This means this problem does not satisfy the
conditions of singularity theorems and they are not applicable in this case.
This is the reason why the test particle has the chance to escape this
curvature singularity.

It is noticable that from (46) we may infer that $A$ and $B^{-1}$
approach to positive zero as $r$ goes to zero. This means 
at $r=0$ the common convention of $t$ as time coordinate
and $r$ as space coordinate remain valid.  This is in agreement
with our initial assumption that the point particle is located
at $r=0$ for ever.\\
We may conclude our discussion by expressing that there is 
no trace of those conceptual problems which was raised
by Narlikar in this form of the metric.

\section{Appendix}

The geodesic equations of this metric are the following, where $p$ is a
parameter describing the trajectory.

\begin{equation}
\frac{d^2r}{dp^2}+\frac{A^{^{\prime }}}{2A}\left( \frac{dr}{dp}\right)
^2-\frac rA\left( \frac{d\theta }{dp}\right) ^2-\frac{r\sin ^2\theta }%
A\left( \frac{d\varphi }{dp}\right) ^2+\frac{B^{^{\prime }}}{2A}\left( \frac{%
dt}{dp}\right) ^2=0
\end{equation}\label{71}

\begin{equation}
\frac{d^2\theta }{dp^2}+\frac 2r\frac{d\theta }{dp}\frac{dr}{dp}-\frac{\sin
^2\theta }2\left( \frac{d\varphi }{dp}\right) ^2=0
\end{equation}\label{72}

\begin{equation}
\frac{d^2\varphi }{dp^2}+\frac 2r\frac{d\varphi }{dp}\frac{dr}{dp}+2\cot
\theta \;\frac{d\theta }{dp}=0
\end{equation}\label{73}

\begin{equation}
\frac{d^2t}{dp^2}+\frac{B^{^{\prime }}}B\frac{dr}{dp}\frac{dt}{dp}=0
\end{equation}\label{74}
Since the field is isotropic, we may consider the orbit of the test particle
to be confined to the equatorial plane,that is $\theta =\frac \pi 2$. The
equation (72) immediately is satisfied and we can forget about $\theta $ as
a dynamical variable. Then equation (73) gives:

\begin{equation}
\frac{d^2\varphi }{dp^2}+\frac 2r\frac{d\varphi }{dp}\frac{dr}{dp}%
=0\;\;\;\;or\;\;\;\;\frac d{dp}\left[ \ln \left( \frac{d\varphi }{dp}\right)
+\ln r^2\right] =0
\end{equation}\label{75}
Integrating (75) with respect to $p$ leads to:

\begin{equation}
\frac{d\varphi }{dp}=\frac J{r^2}
\end{equation}\label{76}
where $J$ is a constant. Dividing (74) by $\frac{dt}{dp}$ gives

\begin{equation}
\frac d{dp}\left[ \ln \left( \frac{dt}{dp}\right) +\ln B\right]
=0\;\;\;\;or\;\;\;\;\frac{dt}{dp}=\frac 1B
\end{equation}\label{77}
The related constant of integration has been absorbed into the definition of 
$p$. Inserting (76) into (71) we obtain

\begin{equation}
\frac{d^2r}{dp^2}+\frac{A^{^{\prime }}}{2A}\left( \frac{dr}{dp}\right) ^2-%
\frac{J^2}{r^3A}+\frac{B^{^{\prime }}}{2AB^2}=0
\end{equation}\label{78}

or

\begin{equation}
\frac d{dp}\left[ A(r)\left( \frac{dr}{dp}\right) ^2+\frac{J^2}{r^2}-\frac
1B\right] =0
\end{equation}\label{79}

By integrating (79) with respect to $p$ we get

\begin{equation}
A(r)\left( \frac{dr}{dp}\right) ^2+\frac{J^2}{r^2}-\frac 1B=-E
\end{equation}\label{80}
where E is a constant of integration and if the test particle is going to be
moveless at infinity we must have $E=1$. Equation (80) for a geodesic with $%
d\varphi =0$ i.e. $J=0$ becomes

\begin{equation}
A(r)\left( \frac{dr}{dp}\right) ^2-\frac 1B=-1
\end{equation}\label{81}
Inserting (81) in (78) with $J=0$ gives

\begin{equation}
\frac{d^2r}{dp^2}+\frac 1{2AB}\left( \frac{A^{^{\prime }}}A+\frac{%
B^{^{\prime }}}B\right) -\frac{A^{^{\prime }}}{2A^2}=0
\end{equation}\label{82}
and inserting (39) in (82) gives

\begin{equation}
\frac{d^2r}{dp^2}+\frac 12\frac d{dr}\left( \frac 1A\right) +\frac{2GM\delta
(r)}{c^2r\sqrt{ABB(0)}}=0
\end{equation}\label{83}
and at last using (47) and (50) in (83) yields the final conclusion

\begin{equation}
\frac{d^2r}{dp^2} =-\frac{GM}{c^2r^2}-\frac{8G^2M^2}{c^4r^3}
(\theta (r)-1)+(\frac{4G^2M^2}{c^4r^2}-2e)\delta (r)
\end{equation}\label{84}
In the r.h.s of (84) the first term exhibits the common attractive force and
the other  terms represent  point interactions which act at $r=0$%
. Indeed at the origin the third is the dominant repulsive force.

\end{document}